\documentclass[preprint,aps,draft]{revtex4}

\usepackage{epsf}

\begin{document}

\title{Trapping Long-Lifetime Excitons in a Two-Dimensional Harmonic Potential}

\author{D.W. Snoke}
\email{snoke@pitt.edu}
\author{Y. Liu}
\author{Z. V\"or\"os}
\affiliation{Department of Physics and Astronomy, University of Pittsburgh,
3941 O'Hara St., Pittsburgh, PA 15260}
\author{L. Pfeiffer}
\author{K. West}
\affiliation{Bell Labs, Lucent Technologies, 700 Mountain Ave., Murray
Hill, NJ 07974}

%\date{June }

\begin{abstract}
We report an important step forward for the goal of unambiguous observation of Bose-Einstein condensation
of excitons in semiconductors. We have demonstrated a system in which excitons live for microseconds,
much longer than their thermalization time, move over distances of hundreds of microns, and can be
trapped in a harmonic potential exactly analous to the traps for atomic condensates. We also report
recent results of a new method for observing evidence of Bose-Einstein condensation, by angular
resolution of the emitted luminescence.
\end{abstract}
\vspace{.5cm}

\maketitle

As far back as the 1960's, theorists have
predicted\cite{mosk,blatt,keld,zimmer,haug,noz,book} that excitons in semiconductors should undergo
Bose-Einstein condensation (BEC) at low temperature. The initial attempts at observing this effect used
bulk semiconductor crystals\cite{peygh,linwolfe,timofeevSi}, but gave ambiguous results.
Some of the barriers to observation of this effect include a competing
phase transition at low density, known as electron-hole liquid \cite{EHL2},
which occurs when the interactions between the excitons are attractive;
short lifetime of excitons in many systems, which prevents them from
ever reaching a thermodynamic equilibrium; and density-dependent
recombination\cite{ohara,denev}, which limits the maximum density in some
systems. 

Over the past decade there
has been concerted effort to observe BEC of excitons in a semiconductor system engineered to overcome
these problems. The system of indirect excitons in coupled quantum wells \cite{yudson,lozo}  
is in many ways an optimal system for observing BEC of excitons, and numerous experimental efforts
\cite{mendez,timofeev,butov,snoke} have been directed toward this goal. In this system, carriers are
trapped in a set of two identical, parallel quantum wells. When voltage is applied normal to the wells, 
the electrons are all drawn to one of the wells, while holes are all drawn to the other well, and
``indirect'' excitons form as bound pairs of spatially separated electrons and holes. This system has
several properties which favor BEC. First, the lifetime of the excitons
is greatly extended, in the present experiments up to 10 $\mu$s, compared to
exciton-phonon scattering times in the range of tens of picoseconds up to nanoseconds, which allows them
to reach thermodynamic equilibrium. Second, since the indirect excitons are aligned dipoles, they have
an overall repulsive interaction, and electron-hole liquid cannot form \cite{lozo}. Last, the quantum
confinement in one direction means that the
exciton gas is two-dimensional, and therefore the total number of excitons
needed for BEC at a given temperature is much lower than in
three dimensions. In typical experiments with bulk crystals, the upper limits of available
laser intensities must be used, while in the quantum well experiments, the needed densities are
easily reached with standard laser systems.

The greatest problem with coupled quantum well structures has been the presence of disorder  \cite{kash},
which arises during the growth process through a number of mechanisms,
such as well width variation, disorder in the alloys used for some of the
epitaxial layers, and impurities. Typical GaAs quantum well structures in
the early 1990's had inhomogeneous luminescence line broadening, which is
a good measure of the disorder energy, of around 5 meV. If the excitons
have kinetic energy less than the disorder energy, the excitons
will mostly be localized in random energy minima; in other words, 
for disorder energy of 5 meV, at temperatures less than around
50 K most of the excitons will be unable to move. If the excitons are localized, one
cannot treat them as a gas to which one can apply equilibrium
thermodynamics of phase transitions.  If the temperature is raised, the excitons can become ionized.
Since the binding energy of excitons in GaAs quantum wells ranges from
5-10 meV, depending on the exact structure\cite{bastard}, a substantial fraction of the excitons will
be ionized if $k_BT$ is large compared to a disorder energy of 5 meV.  Also, the density which is
needed for BEC increases with increasing $T$, so that raising the temperature of the excitons may mean
that the density needed for BEC reaches the range at which the excitons dissociate
due to collisions \cite{crawford}. 

The quality of GaAs quantum well structures, however, has been steadily
improving. One approach which substantially decreases the effect of
disorder, and therefore increases the mobility, is to use wide quantum
wells. Many previous studies have used narrow quantum wells of 50-60 \AA~
GaAs, because the binding energy of the excitons is increased for
narrower wells. The mobility in quantum wells is approximately
proportional to the well width to the sixth power \cite{sakaki}, however,
which strongly favors wider wells. Another factor which favors wider quantum wells is the effect of
well width on the interactions between the excitons. As shown by Lozovik and Berman \cite{lozo}, the
interactions between excitons become completely repulsive for total electron-hole separation of three
times the two-dimensional excitonic Bohr radius, which in GaAs is approximately 65 \AA. For these
reasons, for this study we have used a structure with two 100 \AA GaAs quantum wells, separated by a
thin, 40 \AA~ Al$_{.3}$Ga$_{.7}$As barrier, which has exciton binding energy
at high electric field of 3.5 meV \cite{szy-private}.

A second hurdle to overcome is the effect of a current of free charge carriers passing through the
quantum wells. Because we apply voltage normal to the wells, there is always a current due to drift
of carriers from the surrounding, doped substrate, which tunnel through the outer barriers confining
the electrons and holes in the quantum wells. These carriers will tend to screen out the electron-hole
Coulomb interaction, thus reducing the binding energy, and they will decrease the
mobility of the excitons by random scattering with free carriers. In previous work
\cite{butov-nature,snoke-nature}, the current of free carriers played a major role in the explanation
\cite{snoke-ssc,snoke-simon-prl} of the mysterious effect of luminescence rings. To reduce the current
through our structure, we take employ three strategies. First, we use the highest possible barriers,
Al$_{.45}$Ga$_{.55}$As. Second, we introduce into the barriers a superlattice of 60 \AA~ GaAs wells,
which act to trap free carriers moving across the barriers, following the approach of Dremin et
al.\cite{timofeev}. Last, the excitons are created using a laser with wavelength nearly resonant with
the single-well exciton energy. All of these approaches lead to a strong reduction in the current; for
these samples the typical dark current is less than 1 $\mu$A/cm$^2$. 

All of the above design features lead to a structure in which the indirect excitons have low disorder
and high mobility.  
Figure 1 shows a time-integrated image of the indirect exciton luminescence, following a short (200 fs)
laser pulse resonant with the single-well exciton energy, for the coupled quantum well system described
above, immersed in liquid helium at $T = 2$ K. This image was taken by projecting an image of the sample
onto the entrance slit of an imaging spectrometer with a CCD camera on the back focal plane, so
that the horizontal axis corresponds to the position of the sample while the vertical axis corresponds
to the luminescence photon energy. Although the laser is focused to a tight spot of 30 $\mu$m, the
exciton luminescence is spread over hundreds of microns, due to the motion of the excitons. The bright,
spectrally broad but spatially narrow line is luminescence from the doped GaAs substrate and capping
layer, which are necessarily also excited by the laser. The substrate luminescence has short lifetime and
therefore does not spread out spatially. The exact energy of the indirect excitons depends on the
applied electric field, according to the quantum confined Stark effect \cite{QCSE}. In these
experiments, the indirect exciton line is shifted by high electric field to an energy below the bulk
GaAs band gap.

As seen in Figure 1, the indirect exciton energy is shifted upward in the
center, where the density is highest, and is lower for locations away from the center, falling back to
the unrenormalized value as the density of the excitons drops. This is consistent with the expected
mean-field blue shift of bosons with repulsive interaction potential. As shown in Figure 2, the size of
the exciton cloud becomes larger at higher densities, approximately proportional to the square root of
the laser power, which also indicates that there is a pressure in the exciton gas due to the repulsive
interactions between the excitons.

Figure 3 shows time-resolved luminescence for various distances from the excitation spot. The rise time
of the luminescence is longer for spots further away from the excitation spot, consistent with the
picture that the excitons move out from the excitation spot without becoming localized. The lifetime of
the excitons is 6.5 $\mu$s, in agreement with the calculated \cite{szy-private} enhancement of the
oscillator strength of the indirect excitons. In all cases the period between laser pulses was 4 $\mu$s,
which means that some of the excitons created by each laser pulse will remain at the time of the
succeeding pulse.  Unlike previous experiments
\cite{butov-nature,snoke-nature}, there are no intermediate dark states; the exciton cloud moves
continuously from the laser spot outward.

We conclude that the excitons are delocalized as a free gas, traveling distances of hundreds of microns.
This is an essential point, because the free motion of the excitons over long distances must be
established if we are to believe that we observe a thermodynamic phase transition and not localized
excitons. 

BEC is not possible in two dimensions unless there is a confining potential. Therefore, we create a
harmonic potential minimum in the plane of the excitons using the stress method described elsewhere
\cite{negoita}. One advantage of this method is that we can easily switch between a two dimensional,
translationally invariant system and a confined system by removing or applying the external stress. 
Previous experiments \cite{butov,kash} have attempted to observe a Kosterlitz-Thouless superfluid
transition in a translationally invariant system, but the hydrodynamics of a freely expanding superfluid
are poorly understood. Alternatively, some previous work \cite{timofeev} has centered around accumulation
of excitons in local minima created by disorder in the wells, but these minima are very small and their
exact spatial profile is poorly mapped.

In order for the excitons to act as a gas confined in a potential minimum, the diffusion length of the
excitons must be comparable to the equilibrium size of the exciton cloud in the trap. This
is the case in our experiments for these high-mobility excitons. Figure 4(a) shows the exciton
luminescence energy vs. position in the two-dimensional plane for a harmonic potential created by
stress. The excitons were created on one
side of the potential minimum, and they then flowed from the excitation spot in all directions
toward the energy minimum. Although the potential energy is not strictly a harmonic potential, we can
approximate the potential energy at the minimum as a harmonic potential $U(x) =\frac{1}{2}k
x^2$, with an effective spring constant of $k = 31 \pm 3$ eV/cm$^2$.   As seen in this
figure, the excitons move a long distance from the creation spot, flowing distances comparable to the
size of the harmonic potential trap. Again, this shows that the excitons are delocalized and flow freely
in response to the drift force created by the stress gradient. If we move the laser spot to the center
of the trap, the exciton cloud contracts into a small cloud at the bottom of the trap. Figure 4(b) shows
an image made in the same way, but with the laser moved to the center of the trap. Figure 4(c) shows the
profile of the exciton luminescence intensity in the trap for various times after the laser pulse. At
early times the profile is a sum of the distribution left over from the previous laser pulse and the
newly deposited excitons with spatial width approximately 30 $\mu$m. At late times, the exciton
gas maintains a nearly constant size, which shows that the trap acts to confine the exciton
gas. 

The interactions of the particles act to flatten out the external potential, due to the mean-field,
density-dependent blue shift discussed above. We can deduce this effective potential if we assume that at
late times, the spatial profile of the gas is given by $I(x) \propto e^{-U(x)/k_BT}$, where
$U(x)$ is the sum of the external potential and the internal potential due to interactions. Figure 5
shows a plot of
$U(x)/k_BT = -\ln I(x)$. In this case, the effective spring constant, found by fitting the
potential minimum to a harmonic potential, is approximately 1.7 eV/cm$^2$.

In principle, BEC in a harmonic potential minimum implies a spatial condensation in the lowest
eigenstate of the potential. The interactions between the excitons, however, will cause the
condensate to broaden spatially \cite{leggett}, and for strong interactions the size can become
comparable to the classical equilibrium distribution. As shown by Keeling et al.\cite{levitov}, however,
the in-plane momentum distribution of the condensate will differ strongly from that of particles in
excited states. This implies a strongly peaked {\em angular} distribution of the emitted light.
Essentially, if the condensate is in the lowest momentum state, then the in-plane momentum is zero,
and since the light emission process conserves the in-plane $k$-vector, the light emission should
correspond to a diffraction-limited beam emitted normal to the plane in which the excitons are confined.

Figure 6 shows the angular distribution of the luminescence light emission from the quantum well
structure for several different powers. Typically, one thinks of keeping the density constant and
dropping the temperature to approach the BEC phase transition, but as mentioned above, the temperature
of the excitons is not a control parameter, instead being determined by the rate of phonon emission.
Instead, one can approach the BEC phase transition by keeping all conditions the same and increasing the
particle density, e.g. in our case by simply turning up the laser power. The effective temperature of
the excitons should remain constant, assuming the rate of phonon emission is constant. As seen in Figure
5, at low excitation power, the angular distribution is broad, corresponding to the maximum angular
acceptance of our light collection system, while at high exciton density, a central peak appears. In
other words, the light emission is in the form of a beam, which has angular width of 0.014 radians. This
implies a source size of 70 $\mu$m, assuming  diffraction-limited emission, i.e. $\Delta
\theta = 1.22
\lambda/d$. This is smaller than the equilibrium distribution size of 150 $\mu$m, but a reasonable
estimate of the size of the high-density cloud initially created by the laser pulse. Preliminary
time-resolved measurements indicate that the beam-like emission occurs primarily only at very short time
delays after the laser pulse, consistent with the prediction \cite{levitov-private} that the emission
from the excitonic condensate should be superradiant, with very short lifetime.

This behavior is strikingly different from that of a nondegenerate exciton gas. Luminescence emission
from excitons in quantum wells at $T = 2$ K is emitted in all directions. 
At this point, however, we cannot conclude that the angular peak definitely comes from Bose-Einstein
condensation. Several facts appear inconsisent with this interpretation. First, the angular width of the
peak does not vary strongly with density, although the width is larger than our experimental angular
resolution of 0.5 degrees. Theory \cite{levitov} predicts that the width of the peak should vary with
density. Also, the spectrum of the light emitted from the condensate peak should be narrow compared to
the spectrum from the rest of the gas, while we see just one, broad spectral peak. 

To see whether BEC should be expected, we can compare the estimated density of particles to the
theoretical prediction for condensation. The critical number of particles neeed for
condensation in a harmonic potential in two dimensions, in the Stringari-Pitaveskii limit \cite{Daflovo},
is equal to
$$
N_c = \frac{1.6g(k_BT)^2}{(\hbar\omega_0)^2},
$$
where $g$ is the spin degeneracy (equal to 4 for GaAs excitons), $k_B$ is Boltzmann's constant, and
$\omega_0 = \sqrt{k/m}$ is the natural frequency of the harmonic potential. For the effective spring
constant of 31 eV/cm$^2$, and the effective mass of the excitons in the plane of the quantum wells of
$0.14 m_0$, where
$m_0$ is the vacuum electron mass, this implies a critical number of excitons at $T= 2$K of $N =
10^6$.  The
spring constant of the trap, however, is renormalized by the interactions, as seen in Figure 5. For the
effective spring constant of 1.7 eV/cm$^2$, the critical number is $2\times 10^7$. If the
temperature of the excitons does not cool all the way to 2 K, this number will be higher.

The critical threshold of 10 mW average power, seen in Figure 5, corresponds to
$5\times 10^{10}$ photons per pulse delivered to the sample, accounting for the efficiency of our
optical system and 30\% of the light reflected from the surface of the GaAs. The absorption
length in GaAs for the laser wavelength of 797 nm is approximately 100 $\mu$m, while the total quantum
well thickness is 240
\AA $= .024 \mu$m, which means that the fraction of photons absorbed in the quantum well structure is
approximately
$(.024)/(100) = 2.4\times 10^{-4}$. This implies approximately $1.2\times 10^7$ absorbed photons. This is
comparable to, but below, the critical number for BEC in the trap estimated above for $T= 2$ K. 

One explanation for
the angular peak is that lasing actually occurs during the intense pump pulse which creates the
excitons. This is not expected, since the rate of photon emission is so slow. Alternatively, one can
take note of the fact that immediately after the laser pulse, the volume of the newly created excitons is
much less than the quasiequilibrium size, and therefore the above equilibrium estimate may not apply.
While the exciton gas at early time has relatively small volume, it may exceed the critical density for
condensation. 

In conclusion, we have at our disposal a new system with highly mobile excitons with long lifetime, which
we can trap in a harmonic potential, and we can observe the emitted light with spatial, spectral,
temporal, and angular resolution. Our estimates indicate that the densities of excitons are
either at or just below the expected critical density for condensation. A beamlike emission is observed,
which is either due to lasing or possibly nonequilibrium condensation of the excitons.

{\bf Acknowledgements}. This work has been supported by the National Science Foundation under grant under
Grant No. DMR-0102457, and by the Department of Energy under Grant No. DE-FG02-99ER45780. We thank S.
Denev and R. Balili for important experimental contributions. M. Szymanska provided useful theoretical
calculations for this work.

\newpage

\begin{figure}
\epsfxsize=.99\hsize 
\epsfbox{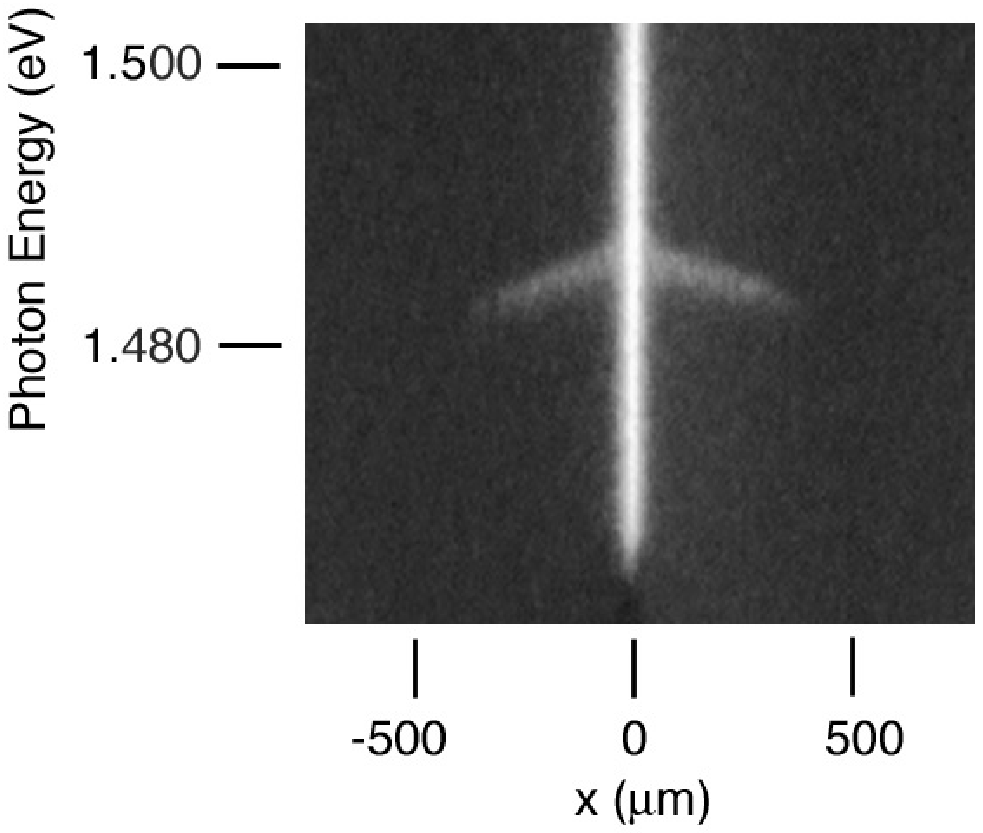} 
\caption{Time-integrated image of the luminescence from the indirect excitons at $T = 2$K, taken using an
imaging spectrometer so that the vertical axis corresponds to the energy of the emitted photons, while
the horizontal axis corresponds to the position in the plane of the quantum wells. The central, spatially
narrow but spectrally broad line corresponds to light emission from the doped GaAs substrate and capping
layers, while the spatially broad but spectrally narrow line corresponds to the emission from the
indirect excitons, of interest in this study.}
\end{figure}

\begin{figure}
\epsfxsize=.99\hsize 
\epsfbox{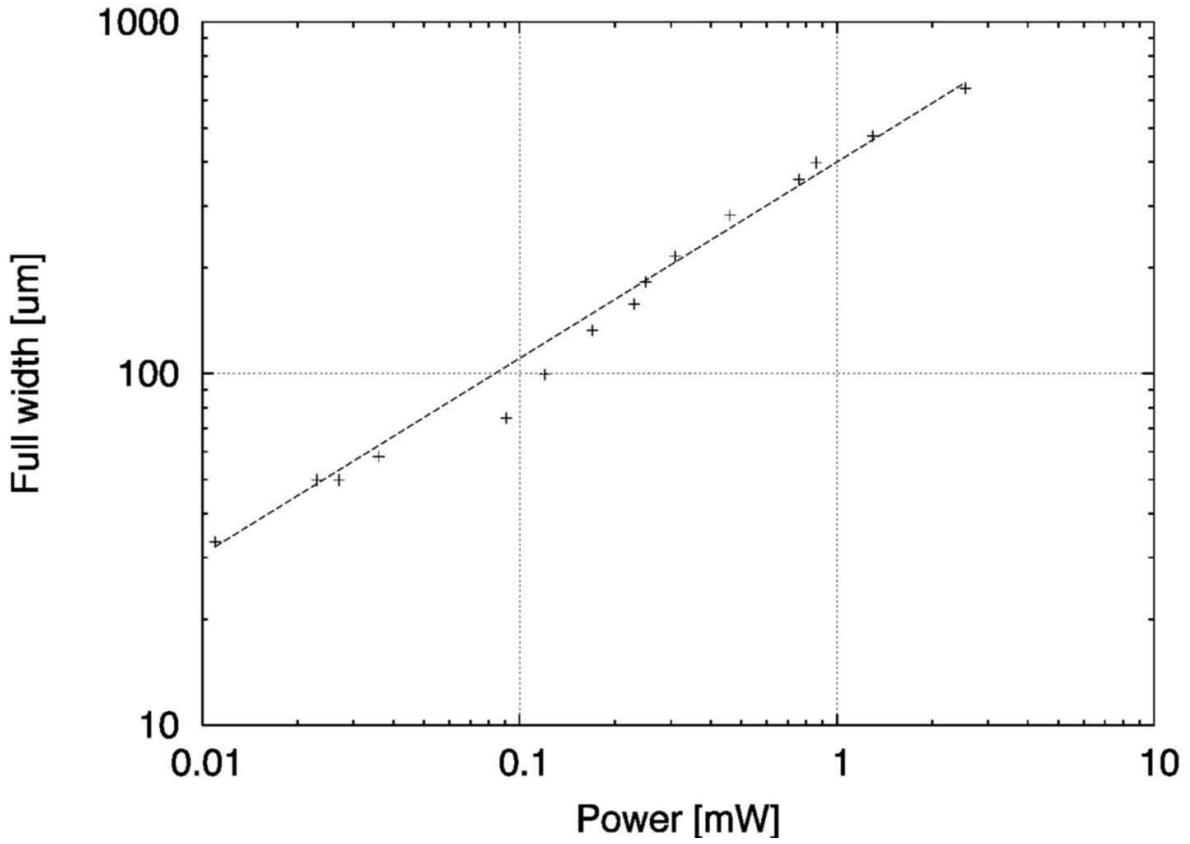} 
\vspace{.5cm}
\caption{Full-width at half-maximum of the exciton cloud as a function of average laser excitation
power.}
\end{figure}

\begin{figure}
\epsfxsize=.99\hsize 
\epsfbox{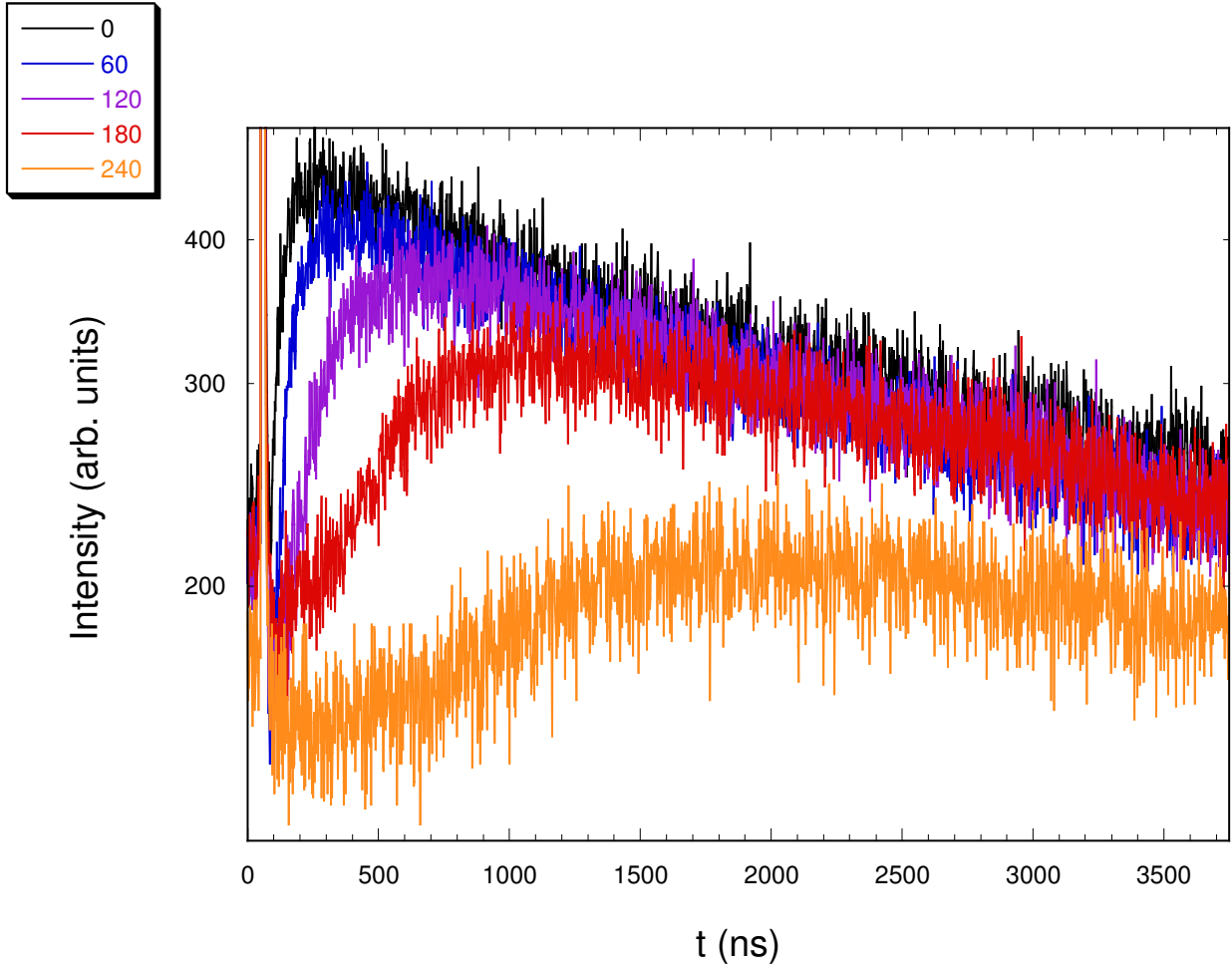} 
\caption{Luminescence intensity as a function of time, for various distances $x$ from the excitation
spot, under the same conditions as Figure 1. The position in the plane is offset by
the same value of $y = 50 \mu$m for all the curves, to reduce the contribution of the light emission from
the substrate and capping layer.}
\end{figure}

\begin{figure}
\epsfxsize=.9\hsize 
\epsfbox{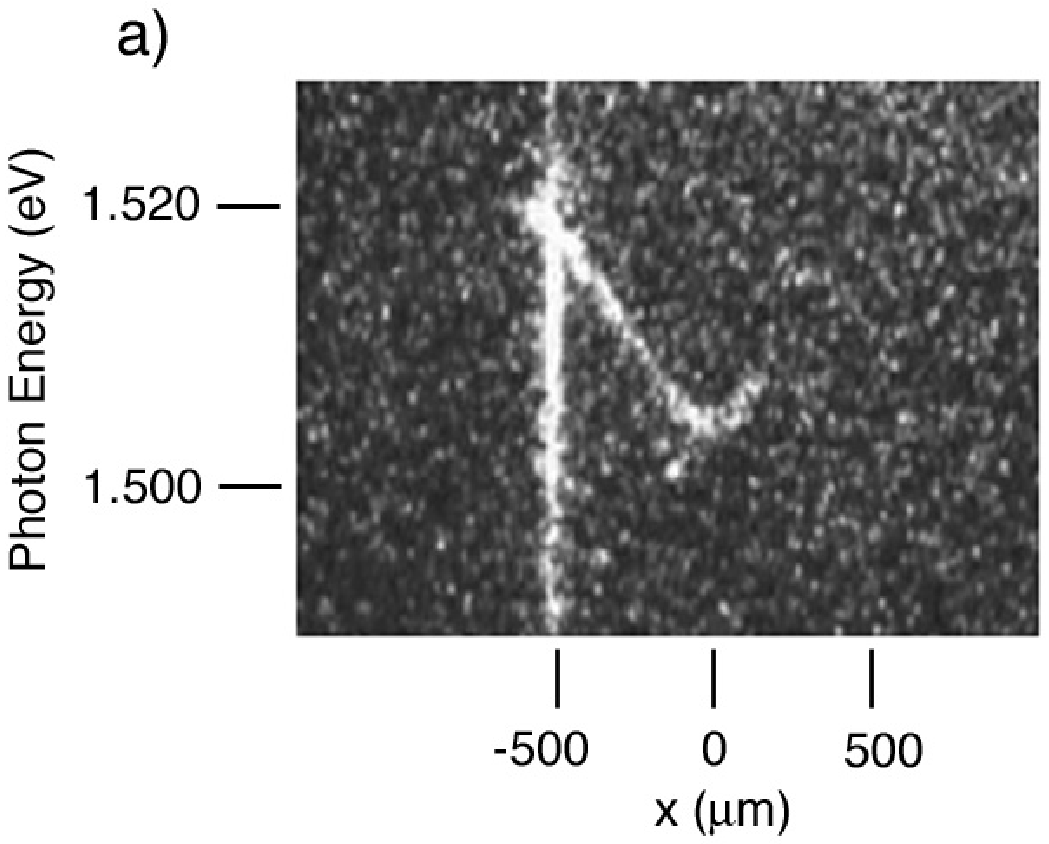} 
\hspace{-1cm}
\epsfxsize=.9\hsize 
\epsfbox{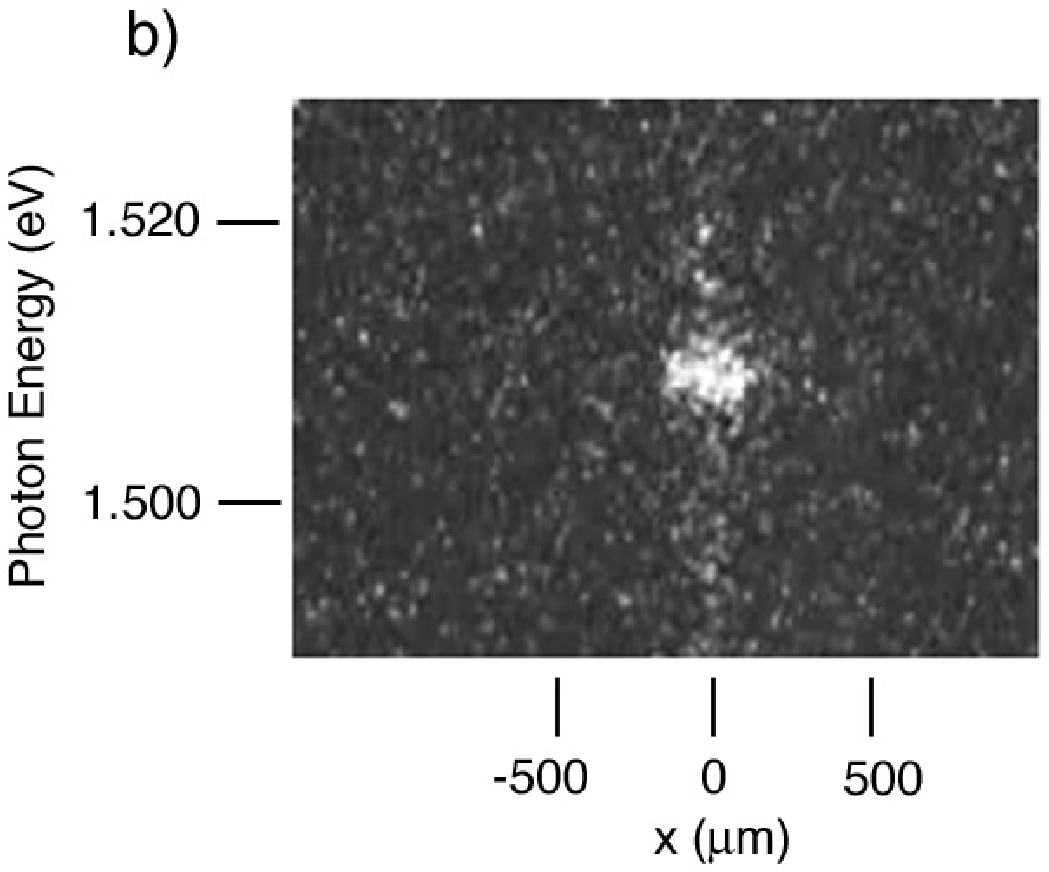} 
\end{figure}
\setcounter{figure}{3}
\begin{figure}
\epsfxsize=.7\hsize 
\epsfbox{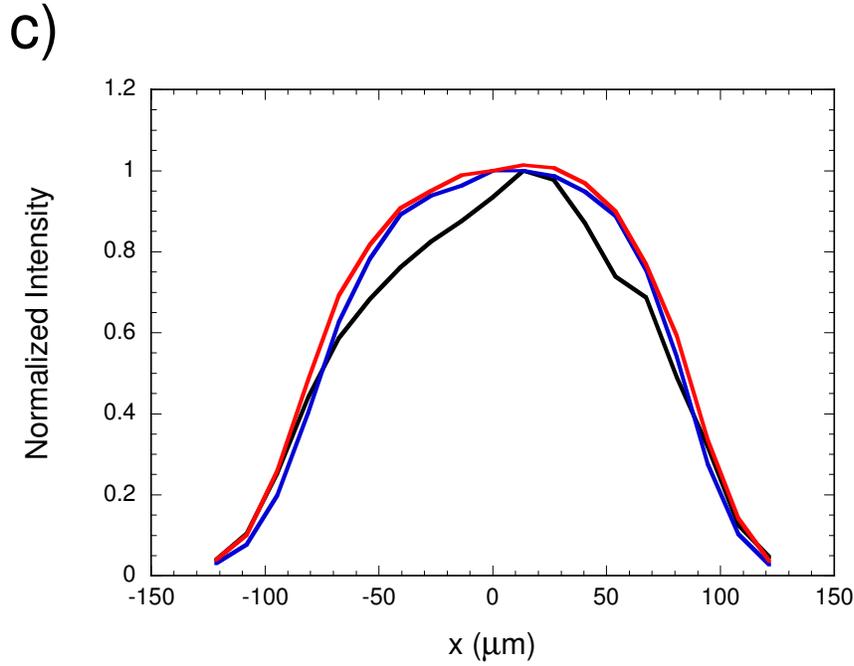} 
\caption{a) Exciton luminescence energy as a function of position, when the laser excitation spot is
located near to a potential minimum created by externally applied stress. The image is recorded in the
same way as Figure 1. b) The same conditions, but the laser spot moved to the center of the potential
minimum. The laser excitation spot has been translated 50 $\mu$m in the perpendicular direction in the
plane of the excitons, to reduce the light from the substrate and capping layer. c) The
intensity of the luminescence as a function of position, for various time delays after the laser pulse,
under the same conditions as (b). Black curve:
$t= 20 $ ns. Blue curve: $t= 1.6$ $\mu$s. Red curve: $t= 3.6$ $\mu$s.} 
\end{figure}

\newpage
\begin{figure}
\epsfxsize=.7\hsize 
\epsfbox{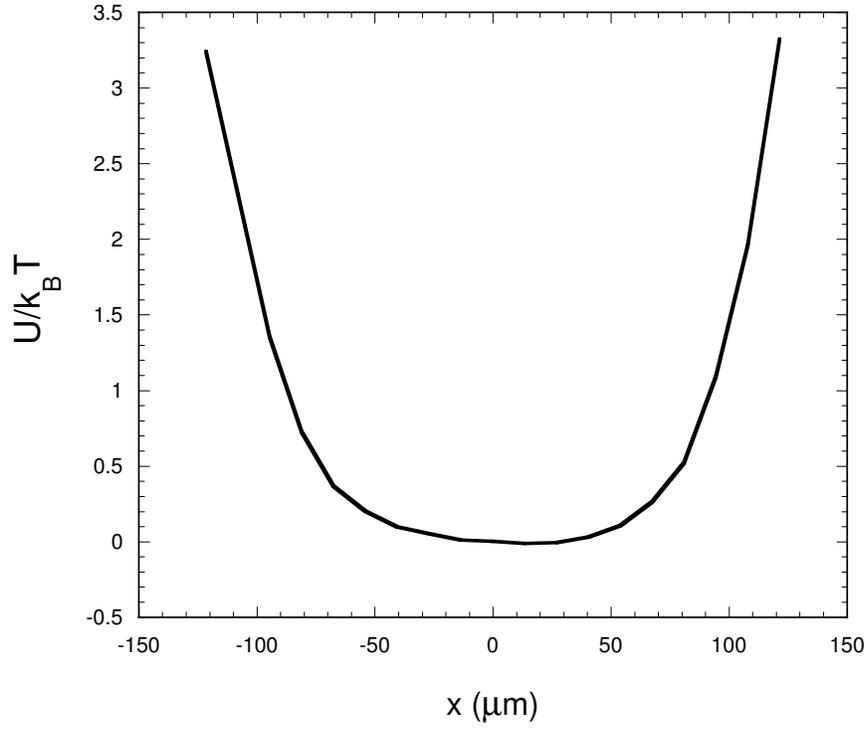} 
\vspace{.5cm}
\caption{The effective potental felt by the excitons, deduced from the intensity distribution in
quasiequilibrium at $t=3.6 \mu$s, shown in Fig. 4(c).}
\end{figure}

\newpage
\begin{figure}
\epsfxsize=.7\hsize 
\epsfbox{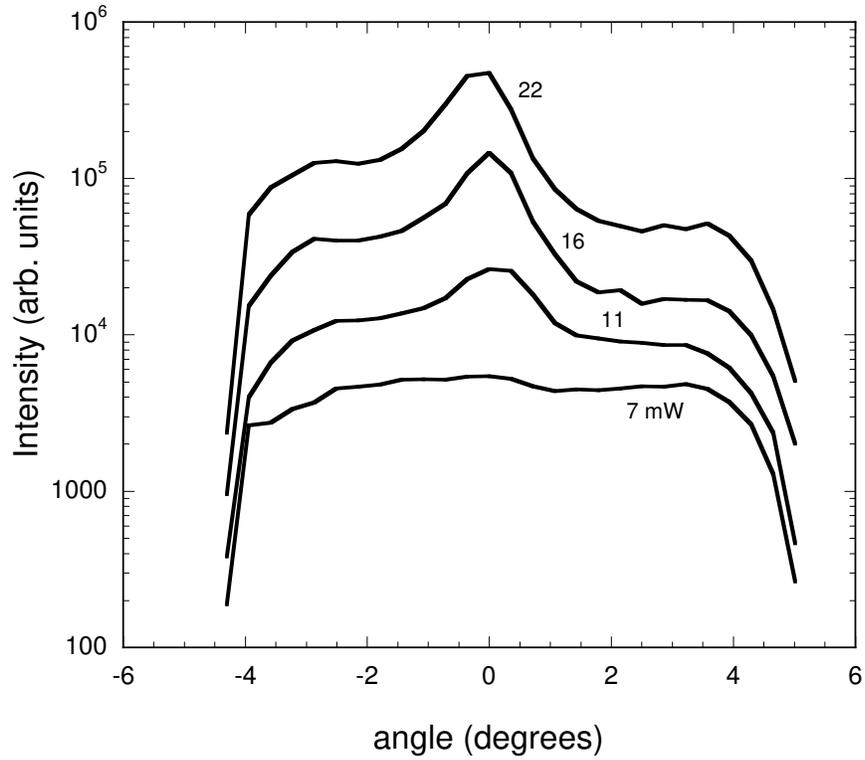} 
\caption{Angular distribution of the light emission from the indirect excitons, for various laser
powers. The angular width of the low-power emission
corresponds to the response of our light collection system to a uniform source; i.e., our system has 9
degrees total angular acceptance. The angular resolution of our system is 0.5 degrees.}
\end{figure}

\end{document}